\documentclass[]{jkas} 
\usepackage{xcolor}

\def\year{2025} 
\def\volume{58} 
\def\issue{1} 
\def\beginpage{1} 
\def\received{---} 
\def\accepted{---} 
\def\published{---} 
\date{Received \received; Accepted \accepted; Published \published}

\title{%
The Low Mass Ratio Overcontact Binary GV Leonis and Its Circumbinary Companion 
}

\author[1,$\star$]{Jae Woo Lee}{0000-0002-5739-9804}
\author[1]{Jang-Ho Park}{0000-0001-9339-4456}
\author[2]{Mi-Hwa Song}{0000-0003-2772-7528}
\author[1,2]{Min-Ji Jeong}{0000-0002-8394-7237}
\author[2]{Chun-Hwey Kim}{0000-0001-8591-4562}

\affil[1]{Korea Astronomy and Space Science Institute, Daejeon 34055, Republic of Korea}
\affil[2]{Department of Astronomy and Space Science, Chungbuk National University, Cheongju 28644, Republic of Korea}

\def\corrauthor{%
Jae Woo Lee, \email{jwlee@kasi.re.kr}
}

\def\runningauthor{%
J. W. Lee et al.
}

\def\runningtitle{%
GV Leonis and Its Circumbinary Companion
}

\def\keywords{%
binaries: close --- binaries: eclipsing --- stars: individual (GV Leonis) --- techniques: photometric --- techniques: spectroscopic 
}

\def\abstracttext{%
Photometric and spectroscopic observations of GV Leo were performed from 2017 to 2024. The light curves show a flat bottom 
at the primary eclipse and the conventional O'Connell effect. The echelle spectra reveal that the effective temperature and 
rotation velocity of the more massive secondary are $T_{\rm eff,2}$ = 5220$\pm$120 K and $v_2 \sin i$ = 223$\pm$40 km s$^{-1}$, 
respectively. Our binary modeling indicates that the program target is a W-subclass contact binary with a mass ratio of 
$q$ = 5.48, an inclination angle of $i$ = 81$^\circ$.68, a temperature difference of ($T_{\rm eff,1}-T_{\rm eff,2}$) = 154 K, 
and a filling factor of $f$ = 36 \%. The light asymmetries were reasonably modeled by a dark starspot on the secondary's 
photosphere. Including our 26 minimum epochs, 84 times of minimum light were used to investigate the orbital period of 
the system. We found that the eclipse times of GV Leo have varied by a sinusoid with a period of 14.9 years and a semi-amplitude 
of 0.0076 days superimposed on a downward parabola. The periodic modulation is interpreted as a light time effect produced by 
an unseen outer tertiary with a minimum mass of 0.26 M$_\odot$, while the parabolic component is thought to be a combination 
of mass transfer (secondary to primary) and angular momentum loss driven by magnetic braking. The circumbinary tertiary would 
have caused the eclipsing pair of GV Leo to evolve into its current short-period contact state by removing angular momentum 
from the primordial widish binary. 
}

\begin{document}
\jkashead 


\section{Introduction}

W UMa-type variable stars are one of the most frequent objects in eclipsing binaries (EBs) \citep{rucinski1969}, and 
are very useful for studying potential stellar mergers such as luminous-red novae and fast-rotating FK Com stars 
\citep{bradstreet1994,tylenda2011,hong2024}. Typically, they contain two solar-type dwarfs with orbital periods of 
$P <$ 1 day, and share a common envelope in physical contact through which mass and energy exchange occurs 
\citep{lucy1968a,lucy1968b,webbink1976,eggleton2012}. Their light curves present similar eclipse depths and continuous 
light changes in the outside-eclipse part. The short-period EBs fall into two subclasses, A and W. In the A-subclass W UMa, 
the more massive star is hotter than its companion and is eclipsed at the primary minimum (Min I), while in the W-subclass 
it is cooler and is obscured during the secondary eclipse \citep{binnendijk1970,binnendijk1977}. 

The contact binaries are considered to originate from initial tidal-locked detached systems, with loss of angular momentum 
via magnetic braking (hereafter AML$_{\rm MB}$), and finally to merge into single stars 
\citep{guinan1988,bradstreet1994,pribulla2006}. In this process, the tertiary objects orbiting the inner close binaries are 
known to play a leading role in forming the initial systems with short periods (e.g., $P <$ 5 days). \citet{pribulla2006} 
reported that many W UMa stars host at least one circumbinary companion. \citet{tokovinin2006} showed that almost all 
main-sequence binaries with $P <$ 3 days reside inside multiple star systems. The existence of the circumbinary objects 
causes a periodic eclipse timing variation to the observer, the so-called light-travel-time \citep[LTT;][]{irwin1952,irwin1959}. 
The eclipse times act as an accurate clock that can be used to detect such outer tertiaries \citep[e.g.,][]{lee2009}. 

\begin{figure*}[!ht]
\centering
\includegraphics[scale=0.7]{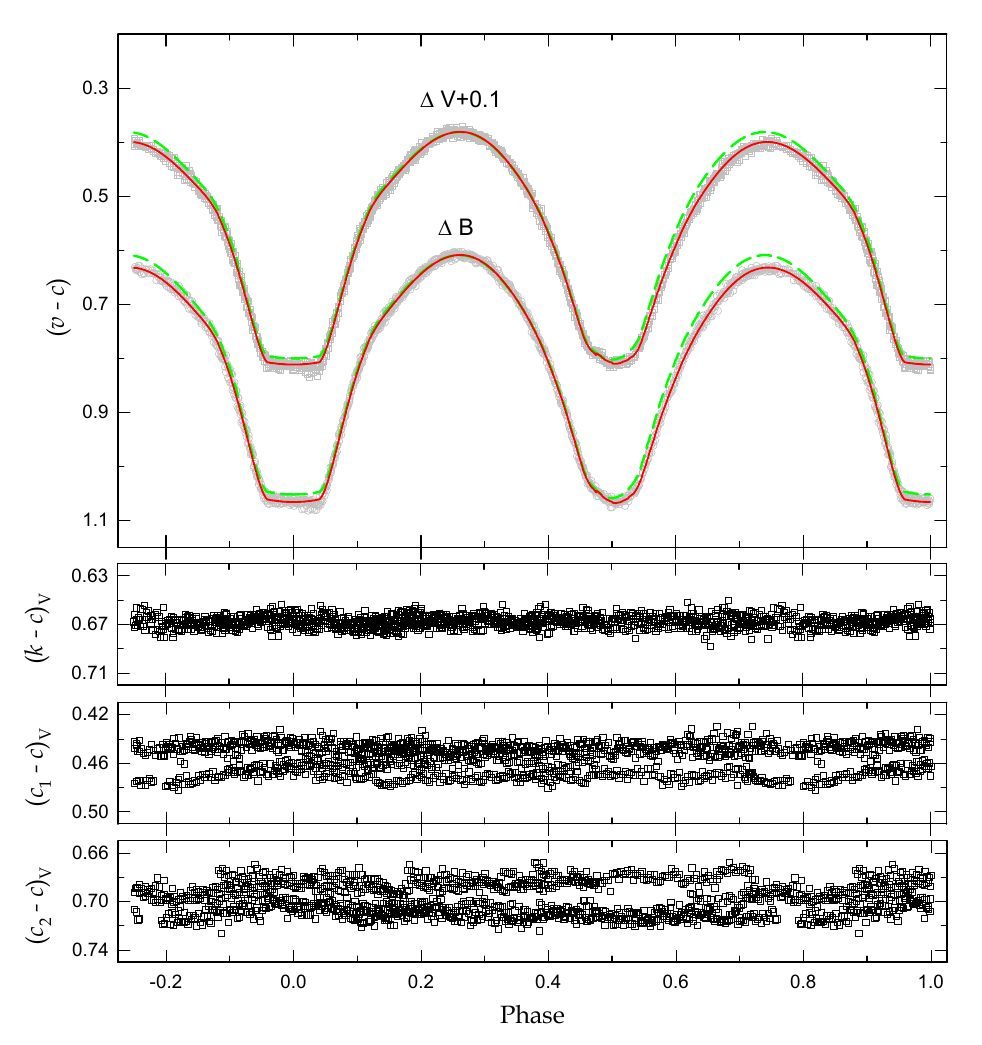}
\caption{Top panel displays $BV$ light curves of GV Leo with the fitted models. The dashed and solid curves represent 
the solutions obtained without and with a dark spot, respectively, listed in Table \ref{Tab1}. Because the two model curves 
partially overlap, much of the unspotted model cannot be seen. The ($k-c$), ($c_1-c$), and ($c_2-c$) differences in 
the $V$ band are shown in the second to bottom panels, respectively, where we can see that the brightness of the $c_1$ and 
$c_2$ stars varied visibly during our observing interval. }
\label{Fig1}
\end{figure*}

This work is concerned with GV Leo (Brh V132, GSC 1419-0091, TYC 1419-91-1, ASAS J101159+1652.5, Gaia DR3 622383646439544320), 
which was announced to be a variable by \citet{bernhard2004}. From his unfiltered light curve, \citet{frank2005} determined 
that the target star is a W UMa EB with a period of $P$ = 0.266727 days. Since then, the multiband light curves for 
the eclipsing variable have been secured by \citet{samec2006} and by \citet{kriwattanawong2013} in the $BVR_{\rm c}I_{\rm c}$ 
and $BVR$ bandpasses, respectively. They solved their own photometric data, using the Wilson-Devinney (W-D) binary code 
\citep{wilson1971,kallrath2022} and applying the starspot model to either of the components \citep{kang1989}. Both results 
indicated that the variable object is a shallow contact binary with a low mass ratio. However, there were notable differences 
in their orbital inclination ($i$), effective temperatures ($T_{\rm eff}$), and luminosities ($L$). \citet{samec2006} presented 
$i$ = 84$^\circ$.7, ($T_{\rm eff,1}-T_{\rm eff,2}$) = 140 K, and $L_2$/($L_{1}$+$L_{2}$){$_{V}$} = 0.790, while \citet{kriwattanawong2013} 
reported $i$ = 76$^\circ$.1, ($T_{\rm eff,1}-T_{\rm eff,2}$) = 494 K, and $L_2$/($L_{1}$+$L_{2}$){$_{V}$} = 0.724. 
These differences may be partly due to the fact that the light curves of \citet{samec2006} showed total eclipses at Min I, 
while those of \citet{kriwattanawong2013} did not.

The orbital period variation for GV Leo was also examined by \citet{samec2006} and \citet{kriwattanawong2013}. The former authors 
suggested that the orbit period is increasing, while the latter reported a secular period decrease. To resolve conflicts in 
the light curve and period studies, we conducted new photometric and first spectroscopic observations and collected all available 
historical data. Through detailed studies of the light curves, echelle spectra, and mid-eclipse timings, we show that GV Leo is 
probably a triple system, comprised of a short-period inner EB and a distant outer companion. In this paper, we refer to 
the hotter component eclipsed at Min I as the primary star (subscript 1) and its companion as the secondary star (subscript 2). 

\begin{figure}[ht]
\centering
\includegraphics[scale=0.6]{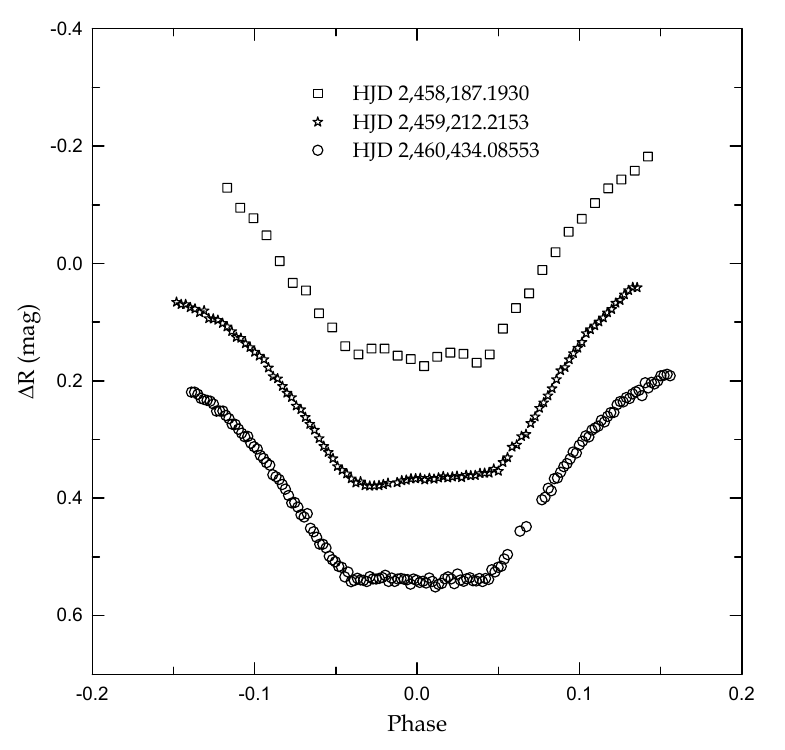}
\caption{Sample $R$-band curves of GV Leo observed during the primary eclipses at SOAO. For clarity, the second and third curves from the top are vertically shifted by 0.2 mag and 0.4 mag, respectively. }
\label{Fig2}
\end{figure}

\begin{figure*}[!ht]
\centering
\includegraphics[scale=0.9]{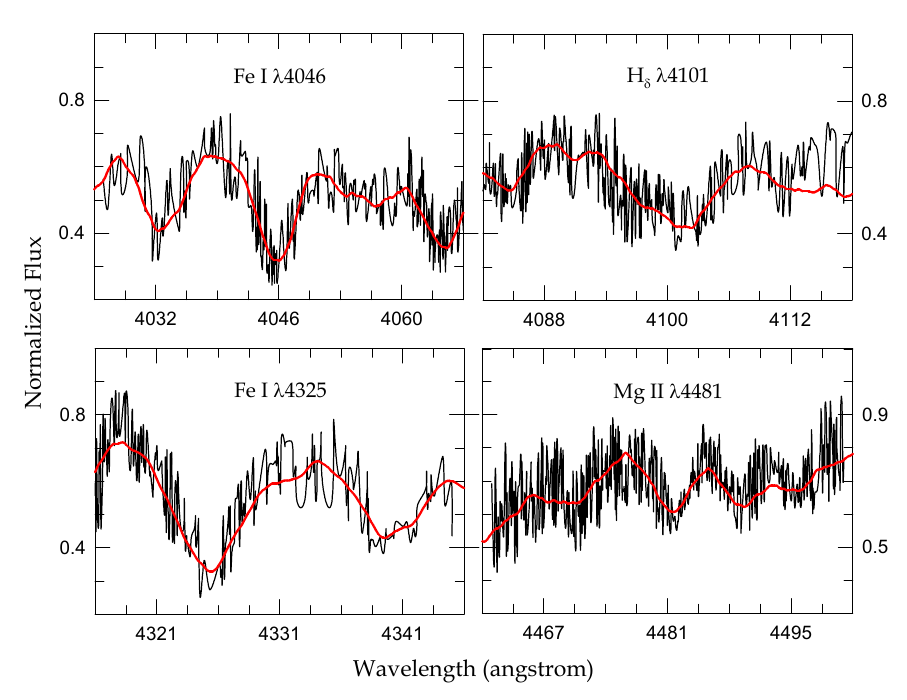}
\caption{Four spectral regions of the more massive secondary star. The black line represents the echelle spectrum observed 
at orbital phase 0.02 (HJD 2,460,031.0691), and the red line is a synthetic spectrum with the best-fit parameters of 
$T_{\rm eff,2}$ = 5220 K and $v_2$$\sin i$ = 223 km s$^{-1}$.}
\label{Fig3}
\end{figure*}

\section{Observations and Data Analysis}

\subsection{Multiband Photometry}

New CCD photometry of GV Leo was performed on January 23, 24, 25, and 27 of 2018, with the 1.0-m telescope at Mt. Lemmon 
Optical Astronomy Observatory (LOAO) in Arizona \citep{han2005}. We secured the multiband light curves using the ARC 4K CCD 
camera and Johnson $BV$ bandpasses. The observational instrument and reduction method employed for the EB system are 
the same as those of HAT-P-12b \citep{lee2012}. Simple aperture photometry using the IRAF package was applied to get 
instrumental magnitudes, whose typical photometric errors are about 0.0017 mag and 0.0015 mag for the $B$ and $V$ bands, 
respectively. To search for a comparison star optimal to GV Leo ($v$), we monitored the EB and nearby stars that were imaged 
simultaneously on the CCD chip. In terms of color, brightness, and constancy in apparent light, TYC 1419-540-1 
(2MASS J10112441+1706216; $c$) and TYC 1419-823-1 (2MASS J10120700+1703157; $k$) were considered suitable comparison and 
check stars, respectively. GSC 1419-0805 ($c_1$) and GSC 1419-1147 ($c_2$), which were used as comparison stars in the observations 
of \citet{samec2006} and \citet{kriwattanawong2013}, seem to be variable stars. 

We obtained 2429 individual points (1213 in $B$ and 1216 in $V$) from the LOAO observations, which are available upon 
request from the first author. The top panel of Figure \ref{Fig1} illustrates the phase-folded $BV$ curves of GV Leo using 
the orbital epoch and period for the dark-spot model provided by our light curve synthesis in Section 3. The $V$-band 
differential magnitudes of ($k-c$), ($c_1-c$), and ($c_2-c$) are displayed in panels (2) to (4), respectively. 
These measurements indicate that our comparison $c$ remained constant in brightness within $\pm$0.005 mag, corresponding to 
the 1$\sigma$-values for both filters, while the other reference stars $c_1$ and $c_2$ changed markedly. 

In addition to the LOAO photometry, we continued observations using the 61-cm telescope and the FLI 4K CCD at SOAO in 
Korea to obtain consistent mid-eclipse times. These observations were made in the $VRI$ bands between 2017 and 2018 and 
in the $R$ band between 2020 and 2024. Figure \ref{Fig2} presents three $R$-band curves at Min I as a sample, 
which show total eclipses but somewhat a tilted flat bottom in the second curve, possibly due to spot activity. Details of 
the SOAO observations were provided in the papers of \citet{park2023,park2024}. 

\subsection{Echelle Spectra}

The effective temperature ($T_{\rm eff}$) of GV Leo has been reported to be in the range of 4800 K to 5300 K. \citet{samec2006} 
and \citet{kriwattanawong2013} assumed the secondary star temperature to be 5000$\pm$300 K and 4850 K (error not given) 
from their photometric data, respectively, while the Gaia DR3 source catalog \citep{gaia2022} listed the EB temperature 
at 5247$_{-12}^{+17}$ K. In contrast, we obtained the intrinsic color index ($B-V$)$\rm_{0,2}$ = +0.79$\pm$0.07 for 
the more massive secondary from both ($B-V$) = $+$0.817$\pm$0.068 at Min I (Samec et al. 2006) and $E(B-V)$ = $A_{\rm V}$/3.1 
= 0.026 \citep{schlafly2011}. This index corresponds to $T_{\rm eff,2}$ = 5300$\pm$200 K \citep{flower1996}, and thus 
to spectral type K0$\pm$2 \citep{pecaut2013}. 

We attempted to understand the atmospheric properties of the GV Leo secondary from high-resolution spectroscopy using the BOES 
spectrograph \citep{kim2007} mounted to the BOAO 1.8-m telescope in Korea. The wavelength range of the BOES is 3600$-$10,200 \rm \AA, 
and we chose a 300 $\mu$m fiber to provide the highest resolution of $R$ = 30,000. Three echelle spectra with exposures of 
450 s each were acquired during Min I on 2023 March 27, when the less massive primary was completely occulted by its companion. 
Before and after the observations, we took spectral images for preprocessing and wavelength correction. The spectroscopic setup, 
data reduction, and spectral analysis was conducted with the same procedure as \citet{lee2023} and \citet{park2023}. 

We applied the $\chi^2$ fitting statistic to four spectral regions of Fe I $\lambda$4046, H$_{\rm \delta}$ $\lambda$4101, 
Fe I $\lambda$4325, and Mg II $\lambda$ 4481 that are appropriate temperature indicators for solar-type stars\footnote
{More information is available on the website: https://ned.ipac.caltech.edu/level5/Gray/frames.html}. This method extracts 
the parameters from a grid search to minimize the difference between observed and model spectra. In this run, our BOES spectra 
were compared to about 50,000 synthetic spectra, covering the ranges ${T_{\rm eff} = 4000-7000}$ K and $v \sin i = 85-250$ km s$^{-1}$. 
The model spectra were interpolated from the ATLAS-9 atmosphere programs of \citet{kurucz1993} by adopting the surface gravity 
of $\log g_2$ = 4.41 (see Section 3), the microturbulent velocity of $\xi$ = 2.0 km s$^{-1}$, and the solar metal abundance. 
Finally, we found the optimal surface temperature and projected rotational velocity for the GV Leo secondary to be 
$T_{\rm eff,2}$ = 5220$\pm$120 K and $v_2 \sin i$ = 223$\pm$40 km s$^{-1}$, respectively. The synthetic spectrum for these parameters 
is plotted in Figure \ref{Fig3}, together with the BOES spectrum observed at phase 0.02. 

\begin{table*}[!ht]
\centering
\caption{GV Leo parameters obtained from the LOAO light curve modeling.}
\begin{tabular}{lccccc}
\toprule
Parameter                                   & \multicolumn{2}{c}{Without Spot}          && \multicolumn{2}{c}{With Spot}             \\ [0.5mm] \cline{2-3} \cline{5-6} \\ [-1.5ex]
                                            & Primary           & Secondary             && Primary           & Secondary             \\ 
\midrule
$T_0$ (HJD)                                 & \multicolumn{2}{c}{2,458,141.85005(52)}   && \multicolumn{2}{c}{2,458,141.849740(32)}  \\
$P$ (day)                                   & \multicolumn{2}{c}{0.2667520(65)}         && \multicolumn{2}{c}{0.2667408(61)}         \\
$q$ (= $M_2/M_1$)                           & \multicolumn{2}{c}{5.403(47)}             && \multicolumn{2}{c}{5.478(15)}             \\
$i$ (deg)                                   & \multicolumn{2}{c}{82.17(72)}             && \multicolumn{2}{c}{81.68(8)}              \\
$T_{\rm eff}$ (K)                           & 5375(130)         & 5220(120)             && 5374(130)         & 5220(120)             \\
$\Omega$                                    & 9.488(82)         & 9.488                 && 9.519(20)         & 9.519                 \\
$\Omega_{\rm in}$$\rm ^a$                   & \multicolumn{2}{c}{9.659}                 && \multicolumn{2}{c}{9.750}                 \\
$f$ (\%)$\rm ^b$                            & \multicolumn{2}{c}{26.7}                  && \multicolumn{2}{c}{36.1}                  \\
$X$, $Y$                                    & 0.648, 0.191      & 0.647, 0.181          && 0.648, 0.191      & 0.647, 0.181          \\
$x_{B}$, $y_{B}$                            & 0.824(28), 0.090  & 0.807(7), 0.054       && 0.660(18), 0.090  & 0.765(4), 0.054       \\
$x_{V}$, $y_{V}$                            & 0.717(25), 0.197  & 0.779(7), 0.170       && 0.570(16), 0.197  & 0.740(4), 0.170       \\
$L_1$/($L_{1}$+$L_{2}$){$_{B}$}             & 0.2147(11)        & 0.7853                && 0.2241(6)         & 0.7759                \\
$L_1$/($L_{1}$+$L_{2}$){$_{V}$}             & 0.2120(11)        & 0.7880                && 0.2197(6)         & 0.7803                \\
$r$ (pole)                                  & 0.2364(22)        & 0.5005(11)            && 0.2386(4)         & 0.5042(3)             \\
$r$ (side)                                  & 0.2471(25)        & 0.5485(17)            && 0.2498(5)         & 0.5539(5)             \\
$r$ (back)                                  & 0.2880(46)        & 0.5729(22)            && 0.2948(9)         & 0.5791(6)             \\
$r$ (volume)$\rm ^c$                        & 0.2582(30)        & 0.5411(17)            && 0.2619(6)         & 0.5461(5)             \\ 
Colatitude (deg)                            & \dots             & \dots                 && \dots             & 57.34(68)             \\
Longitude (deg)                             & \dots             & \dots                 && \dots             & 304.30(55)            \\
Radius (deg)                                & \dots             & \dots                 && \dots             & 17.17(16)             \\
$T$$\rm _{spot}$/$T$$\rm _{local}$          & \dots             & \dots                 && \dots             & 0.937(10)             \\
$\sum W(O-C)^2$                             & \multicolumn{2}{c}{0.0058}                && \multicolumn{2}{c}{0.0030}                \\
\bottomrule
\end{tabular}
\tabnote{
$^{\rm a}$ Potential for the inner critical Roche surface. $^{\rm b}$ Fill-out factor. $^{\rm c}$ Mean volume radius.
}
\label{Tab1}
\end{table*}

\section{Binary Modeling and Fundamental Parameters}

The LOAO observations for GV Leo in Figure \ref{Fig1} show W UMa-like light curves and flat bottoms at primary minima. These 
strongly suggest that the smaller but hotter component is totally obscured by the more massive companion, which implies that 
the program target is a W-subclass W UMa EB. Our multiband light curves indicate that Max I is brighter than Max II by amounts 
of $\sim$0.023 mag and $\sim$0.018 mag in the $BV$ bands, respectively. Further, these maximum light phases are somewhat shifted 
to around 0.26 and 0.74. The brightness disturbances are generally indicative of starspot activity in the photospheres of 
the components \citep[e.g.,][]{kouzuma2019}. 

To get the light curve parameters of GV Leo, we modeled all of the LOAO observations together applying the W-D program. 
In the modeling, we set the surface temperature from our BOES spectral analysis to that (5220$\pm$120 K) of the larger, 
more massive secondary. The albedos of $A_{1,2}$ = 0.5 \citep{rucinski1969} and the gravity-darkening parameters of 
$g_{1,2}$ = 0.32 \citep{lucy1967} were used as standard values for convective dwarfs from the components' temperatures. 
Bolometric ($X$, $Y$) and monochromatic ($x$, $y$) limb-darkening parameters were initialized from the updated logarithmic 
coefficients of \citet{van1993}. Since the observed $v_2 \sin i$ agreed with the computed synchronous rotation 
$v_{\rm 2,sync}$ = $182\pm3$ km s$^{-1}$ from $2 \pi R_{\rm 2}$/$P$, we took the ratios of the rotational and orbital velocities 
for both components to be $F_{1,2}$ = 1.0. 

\begin{figure}[t]
\centering
\includegraphics[scale=0.6]{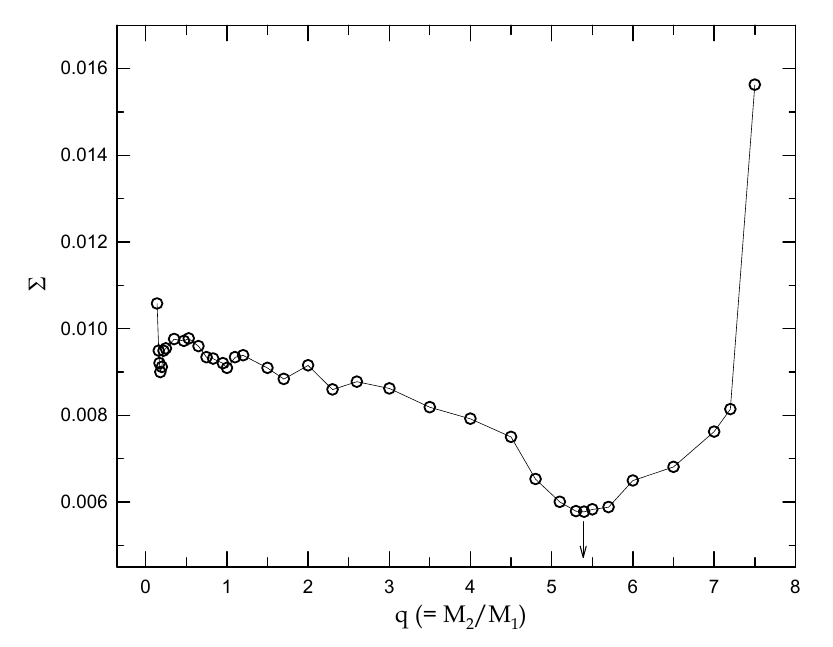}
\caption{Behavior of $\sum$ (the weighted sum of the residuals squared) of GV Leo as a function of mass ratio $q$, showing 
a minimum value at $q$ = 5.40. The circles represent the $q$-search results for each assumed mass ratio.}
\label{Fig4}
\end{figure}

\begin{figure*}[!ht]
\centering
\includegraphics{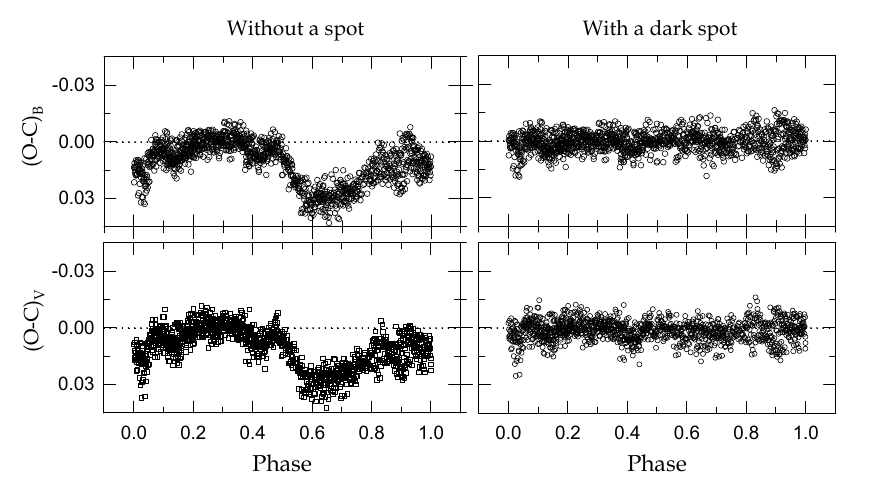}
\caption{Light curve residuals of $B$ and $V$ bands corresponding to the two binary models in columns (2) to (5) of 
Table \ref{Tab1}: without (left panels) and with (right panels) a dark starspot on the secondary component. }
\label{Fig5}
\end{figure*}

The spectroscopic mass ratio $q$ has never been made for GV Leo. Also, although photometric solutions were obtained by 
\citet{samec2006} and \citet{kriwattanawong2013}, their light curves and corresponding parameters did not match each other, 
and the latter authors reported that the eclipsing variable is an A-subclass contact system. Therefore, we performed 
an intensive $q$-search procedure to find an initial $q$ value. The result is presented in Figure \ref{Fig4}, indicating that 
the primary eclipse is an occultation and thus GV Leo is in the W-subgroup of W UMa stars. We solved the LOAO photometric data 
by including the $q$ value as an adjustable parameter. The unspotted solutions are given in columns (2)$-$(3) of Table \ref{Tab1}, 
and the values with errors in parentheses indicate adjustable parameters. The synthetic light curves and their corresponding 
residuals are presented as green dashed lines on the uppermost panel in Figure \ref{Fig1} and as circles on the left panels 
in Figure \ref{Fig5}, respectively. 

As presented in Figures \ref{Fig1} and \ref{Fig5}, the unspotted model does not fit the LOAO data satisfactorily, because 
the light levels at the quadratures are asymmetrical. Thus, we applied possible spot models on the component stars to 
account for the light discrepancy, and present the modeling results in columns (4)$-$(5) of Table \ref{Tab1}. 
The red solid curves in Figure \ref{Fig1} represent the spotted model and the right panels of Figure \ref{Fig5} show 
their corresponding residuals. We can see that the dark starspot on the more massive secondary best matches 
the light asymmetries, resulting in $\sum W(O-C)^2$ being much lower than that of the unspotted solution. In all of 
the light curve syntheses, we considered a third light contribution but the $\ell_{3}$ parameter usually had a negative value. 
On the other hand, we split the LOAO observations into five datasets, solved each of them with the binary modeling code, and 
computed the standard deviations ($\sigma$) for the different values of each parameter. The 1$\sigma$-values are indicated as 
the parameters' errors in Table \ref{Tab1}. 

Our synthesis indicates that GV Leo is a W-subclass overcontact EB with the parameters of $q$ = 5.48, $i$ = 81$^\circ$.68, 
and ($T_{\rm eff,1}-T_{\rm eff,2}$) = 154 K. The secondary's temperature is appropriate for a spectral type between K0 and K1 dwarfs 
and a mass of $M_2$ = 0.87$\pm$0.03 M$_\odot$, based on the updated version of \citet{pecaut2013} on 2022 April 16. 
The absolute parameters for each component presented in Table \ref{Tab2} were calculated from our binary model and 
the $M_2$ value. Bolometric corrections (BCs) were adopted according to the temperature correlation of \citet{torres2010}. 
Using the interstellar extinction of $A_{\rm V}$ = 0.080 \citep{schlafly2011} and $V$ = +11.778$\pm$0.030 at maximum light 
\citep{samec2006}, we derived the geometric distance to GV Leo of 197$\pm$11 pc. Despite the fact that the absolute parameters 
were obtained without double-lined radial velocities, our distance agrees very well with the Gaia DR3 distance of 199$\pm$2 pc, 
calculated from a parallax of 5.027$\pm$0.055 mas \citep{gaia2022}.

\begin{table*}[t]
\centering
\caption{Absolute parameters for GV Leo.}
\begin{tabular}{lcc}
\toprule
Parameter              & Primary             & Secondary              \\
\midrule
$M$ (M$_\odot$)        & 0.16$\pm$0.01       & 0.87$\pm$0.03          \\
$R$ (R$_\odot$)        & 0.46$\pm$0.01       & 0.96$\pm$0.02          \\
$\log$ $g$ (cgs)       & 4.31$\pm$0.02       & 4.41$\pm$0.02          \\
$\rho$ ($\rho_\odot$)  & 1.63$\pm$0.10       & 0.98$\pm$0.06          \\
$L$ (L$_\odot$)        & 0.16$\pm$0.02       & 0.61$\pm$0.06          \\
$M_{\rm bol}$ (mag)    & $+$6.73$\pm$0.11    & $+$5.26$\pm$0.11       \\
BC (mag)               & $-$0.17$\pm$0.04    & $-$0.22$\pm$0.04       \\
$M_{\rm V}$ (mag)      & $+$6.90$\pm$0.12    & $+$5.48$\pm$0.12       \\
Distance (pc)          & \multicolumn{2}{c}{197$\pm$11}               \\
\bottomrule
\end{tabular}
\label{Tab2}
\end{table*}

\section{Eclipse Timing Variation}

Twenty-six minimum epochs and their errors were measured from our LOAO and SOAO observations using the method of \citet{kwee1956}. 
These are given in Table \ref{TabA1}, together with other available CCD epochs. As shown in this table, two secondary minima 
(HJD 2,452,763.3966 and HJD 2,452,764.4639) were newly derived from the unfiltered observations of \citet{frank2005} and 
one primary minimum (HJD 2,453,715.2308) from the public archive of the All Sky Automated Survey \citep[ASAS;][]{pojmanski1997}. 
Because the ASAS data (HJD 2,452,622.83$-$2,454,573.56) are not time-series observations, we calculated the minimum epoch by 
fitting only the reference epoch $T_0$ and period $P$ among the light curve parameters of Table \ref{Tab1} using the W-D program.  

\begin{figure*}[!ht]
\centering
\includegraphics[scale=0.9]{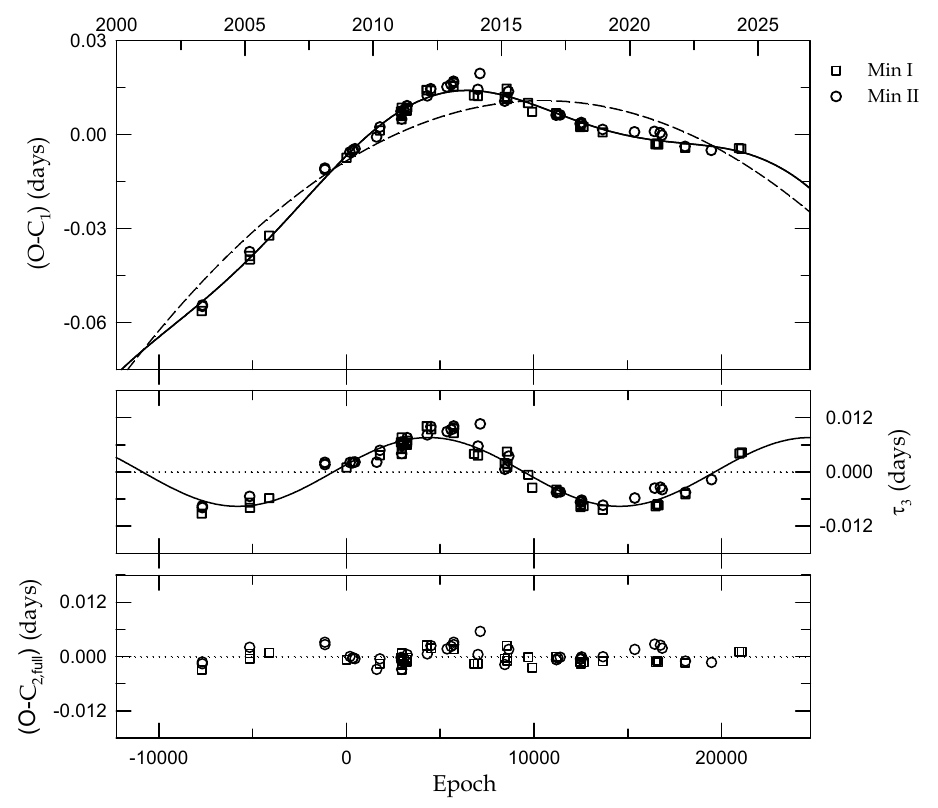}
\caption{Eclipse timing diagram of GV Leo constructed with the linear ephemeris (1). In the top panel, the solid and dashed curves 
represent the full non-linear contribution and just the parabolic term of the quadratic {\it plus} LTT ephemeris, respectively. 
The middle panel refers to the LTT orbit ($\tau_3$) and the bottom panel shows the residuals from the complete $C_2$ ephemeris. }
\label{Fig6}
\end{figure*}

The period change of GV Leo was studied for the first time by \citet{samec2006}. They suggested that there was an upward 
parabolic change in eclipse timings, implying a secular period increase. In contrast, \citet{kriwattanawong2013} reported 
that the period continuously decreased at a rate of $-4.95 \times 10^{-7}$ day year$^{-1}$ from a quadratic least-squares fit, 
resulting from mass exchange between the EB components. As a starting point for our analysis, we fit all minimum epochs 
to obtain the mean orbital ephemeris of GV Leo, as follows: 
\begin{equation}
 C_1 = \mbox{HJD}~ 2,454,814.974156(32) + 0.2667259611(21)E. 
\end{equation}
The $O$--$C_1$ residuals computed by equation (1) are represented in the top panel of Figure \ref{Fig6}. The eclipse timing variation 
(ETV) appears to be due to more than one cause, rather than a simple parabolic change as suggested by previous researchers.

After some trials, we found that the ETV of GV Leo is best represented as a combination of a parabola and a sinusoid. 
The oscillation was provisionally considered as an LTT produced by a tertiary companion orbiting the inner eclipsing pair. 
Thus, we introduced the timing residuals into the following ephemeris:
\begin{eqnarray}
C_2 = T_0 + PE + AE^2 + \tau_{3}. 
\end{eqnarray}
Here, $\tau_{3}$ is the LTT including $a_{\rm b}\sin i_3$, $e_{\rm b}$, $\omega_{\rm b}$, $n_{\rm b}$, and $T_{\rm b}$ 
\citep{irwin1952,irwin1959}. The Levenberg-Marquardt procedure \citep{press1992} was employed to solve equation (2), 
the results of which are detailed in Table \ref{Tab3} and illustrated in Figure \ref{Fig6}. 
The $O$--$C_{\rm 2, full}$ residuals from the full contribution are given in column (4) of Table \ref{TabA1} and presented 
in the lowermost panel of Figure \ref{Fig6}. Here, the minimum epochs agree satisfactorily with our quadratic {\it plus} LTT 
ephemeris. The LTT orbit has a cycle length of $P_{\rm b}$ = 14.9 years and a semi-amplitude of $K_{\rm b}$ = 0.0076 days. 
The mass function of the tertiary component is $f(M_{3})$ = 0.010 M$_\odot$, and its minimum mass is $M_{3}$ = 0.26 M$_\odot$. 

\begin{table*}[!ht]
\centering
\caption{Parameters for the quadratic {\it plus} LTT ephemeris of GV Leo.}
\begin{tabular}{lcc}
\toprule
Parameter                  & Values                                   & Unit                  \\
\midrule
$T_0$                      & 2,454,814.96584$\pm$0.00057              & HJD                   \\
$P$                        & 0.266729629$\pm$0.000000059              & day                   \\
$A$                        & $-$(1.751$\pm$0.040)$\times 10^{-10}$    & day                   \\
$a_{\rm b}\sin i_{3}$      & 1.32$\pm$0.13                            & au                    \\
$e_{\rm b}$                & 0.00$\pm$0.21                            &                       \\
$\omega_{\rm b}$           & 7.9$\pm$6.6                              & deg                   \\
$n_{\rm b}$                & 0.0664$\pm$0.0023                        & deg day$^{-1}$        \\
$T_{\rm b}$                & 2,454,745$\pm$99                         & HJD                   \\
$P_{\rm b}$                & 14.85$\pm$0.51                           & year                  \\
$K_{\rm b}$                & 0.00761$\pm$0.00076                      & day                   \\
$f(M_{3})$                 & 0.0104$\pm$0.0011                        & M$_\odot$             \\
$M_{3} \sin i_{3}$         & 0.258$\pm$0.015                          & M$_\odot$             \\
$a_{3} \sin i_{3}$         & 5.26$\pm$0.15                            & au                    \\
$e_3$                      & 0.00$\pm$0.21                            &                       \\
$\omega_{3}$               & 187.9$\pm$6.6                            & deg                   \\
$P_3$                      & 14.85$\pm$0.51                           & year                  \\
rms scatter                & 0.0016                                   & day                   \\
\bottomrule
\end{tabular}
\label{Tab3}
\end{table*}

\begin{table*}[!ht]
\centering
\caption{Model parameters for possible magnetic activity of GV Leo.}
\begin{tabular}{lccc}
\toprule
Parameter                 & Primary                 & Secondary               & Unit                   \\
\midrule
$\Delta P$                &  0.203                  &  0.203                  &  s                     \\
$\Delta P/P$              &  $8.82\times10^{-6}$    &  $8.82\times10^{-6}$    &                        \\
$\Delta Q$                &  ${4.68\times10^{48}}$  &  ${2.54\times10^{49}}$  &  g cm$^2$              \\
$\Delta J$                &  ${3.33\times10^{46}}$  &  ${1.08\times10^{47}}$  &  g cm$^{2}$ s$^{-1}$   \\
$I_{\rm s}$               &  ${2.18\times10^{52}}$  &  ${5.15\times10^{53}}$  &  g cm$^{2}$            \\
$\Delta \Omega$           &  ${1.53\times10^{-6}}$  &  ${2.10\times10^{-7}}$  &  s$^{-1}$              \\
$\Delta \Omega / \Omega$  &  ${5.62\times10^{-3}}$  &  ${7.72\times10^{-4}}$  &                        \\
$\Delta E$                &  ${1.02\times10^{41}}$  &  ${4.56\times10^{40}}$  &  erg                   \\
$\Delta L_{\rm rms}$      &  ${6.84\times10^{32}}$  &  ${3.06\times10^{32}}$  &  erg s$^{-1}$          \\
                          &  0.175                  &  0.078                  &  L$_\odot$             \\
                          &  1.097                  &  0.128                  &  $L_{1,2}$             \\
$\Delta m_{\rm rms}$      &  $\pm$0.223             &  $\pm$0.105             &  mag                   \\
$B$                       &  20.7                   &  12.4                   &  kG                    \\
\bottomrule
\end{tabular}
\label{Tab4}
\end{table*}

The LTT hypothesis is not be the only possible explanation for the sinusoidal variation. In solar-type contact binaries, 
it is alternatively possible that the period change comes from a modulation in magnetic activity \citep{applegate1992,lanza1998}. 
To describe the period modulation of $\Delta P/P \sim 10^{-5}$, the Applegate mechanism generally requires that 
the magnetically active star should rotate differentially at $\Delta \Omega/\Omega_* \simeq 0.01$ and be variable at 
$\Delta L/L_* \simeq 0.1$. The Applegate parameters for each component were computed using the LTT period ($P_{\rm b}$) and 
amplitude ($K_{\rm b}$) of GV Leo. They are given in Table \ref{Tab4}, wherein the binary components of $L_1$ = 0.16 L$_\odot$ 
and $L_2$ = 0.61 L$_\odot$ show rms variations of $L_{\rm rms,1}$ = 0.18 L$_\odot$ and $L_{\rm rms,2}$ = 0.08 L$_\odot$ in 
the same order. Moreover, the variations ($\Delta Q$) of the gravitational quadrupole moment are much lower compared to 
typical values of $10^{51}-10^{52}$ for W UMa binaries \citep{lanza1999}. These imply that this type of mechanism does 
not adequately explain the timing variation observed in GV Leo. Alternatively, the sinusoidal oscillation can be produced by 
apsidal motion in eccentric binaries. However, our binary model indicates that the eclipsing components of GV Leo are in 
a circular-orbit overcontact configuration, and the timing residuals from both eclipses (Min I and II) are consistent with 
each other. Thus, at present, there is no other alternative but the LTT due to a unseen circumbinary companion. 

The quadratic coefficient $A$ in equation (2) represents the secular component of the ETV, and its negative value indicates 
a period decrease of (d$P$/d$t$)$_{\rm obs}$ = $-4.8\times$10$^{-7}$ day year$^{-1}$. In contact binaries, such a change 
can be considered as the secondary to primary mass transfer and/or AML$_{\rm MB}$. 
Under conservative assumptions, the observed (d$P$/d$t$)$_{\rm obs}$ gives a mass transfer rate of 1.2$\times$10$^{-7}$ M$_\odot$ year$^{-1}$, 
which is 5.5 times larger than the predicted rate of $M_{2}/ \tau_{\rm th}$ = 2.2$\times$10$^{-8}$ M$_\odot$ year$^{-1}$ on 
a thermal time scale of $\tau_{\rm th}$ = $(GM_{\rm 2}^2)/(R_{\rm 2}L_{\rm 2})$ = 4.0$\times 10^{7}$ years \citep{paczynski1971}. 
Moreover, the rate of (d$P$/d$t$)$_{\rm obs}$ is 3.6 times larger than the theoretical AML$_{\rm MB}$ rate of (d$P$/d$t$)$_{\rm AML}$ 
= $-1.3\times$10$^{-7}$ day year$^{-1}$, calculated for the gyration constant $k^2$ = 0.1 in the approximate expression of 
\citet{guinan1988}. Thus, the period decrease in GV Leo may be the result of a combination of these two causes. 


\section{Summary and Discussion}

In this work, we have presented photometric and spectroscopic observations of GV Leo, and analyzed them in detail. 
The light curves indicate that the primary minima display total eclipses, and the light levels at the quadratures are 
asymmetrical. From the echelle spectra, the effective temperature and rotation velocity of the GV Leo secondary were measured 
to be $T_{\rm eff,2}$ = 5220$\pm$120 K and $v_2 \sin i$ = 223$\pm$40 km s$^{-1}$, respectively. 
Our binary model represents that the eclipsing pair is a totally eclipsing W-subclass contact system with a moderate filling 
factor of $f$ = ($\Omega_{\rm in}$--$\Omega$)/($\Omega_{\rm in}$--$\Omega_{\rm out}$)$\times$100 = 36 \%. Here, $\Omega_{\rm in}$ 
and $\Omega_{\rm out}$ are the potentials of the inner and outer critical Roche lobes, and $\Omega$ is the potential corresponding 
to the common envelope of GV Leo. 
The light asymmetries were well matched to a dark starspot model on the secondary component. The fundamental parameters for 
GV Leo were used to locate each component on the $M-R$, $M-L$, and Hertzsprung-Russell (H-R) diagrams \citep[cf.,][]{lee2014}. 
The more massive secondary resides inside the main-sequence band, while the hotter companion, with a very low mass of 0.16 M$_\odot$, 
is oversized and overluminous for its mass, but its location in the H-R diagram is to the left of this band. Such a feature may 
be caused by a significant energy flow from the secondary \citep{lucy1968a,li2008}. 

Detailed analyses of the eclipse timing diagram showed that the orbital period experiences a 15-year oscillation superimposed 
on a downward parabola. In principle, the periodic variation can be produced by three physical causes, but both a magnetic 
activity cycle and apsidal motion are ruled out. Therefore, the observed period modulation of GV Leo most likely comes from 
the LTT via an outer circumbinary object with a projected mass of $M_{3} \sin i_{3}$ = 0.26 M$_\odot$ in a near-circular orbit. 
The third-body mass depends on its orbital inclination with respect to the inner EB, with a smaller $i_{3}$ resulting in 
a larger $M_{3}$ mass. Thus, the outer companion has masses of 0.26 M$_\odot$, 0.31 M$_\odot$, and 0.61 M$_\odot$, respectively, 
for inclinations $i_{3}$ = 90 deg, 60 deg, and 30 deg. Assuming that the circumbinary companion is a normal main sequence and 
its orbit is coplanar with that of the close pair of GV Leo ($i_{3}$ = 81.68 deg), the mass and radius of the tertiary are 
$M_3 \simeq$ 0.26 R$_\odot$ and $R_3 \simeq$ 0.30 R$_\odot$ \citep{pecaut2013}. Then, the circumbinary component has 
a spectral type of M3$-$4V and a bolometric luminosity of $L_3 \simeq$ 0.01 L$_\odot$, which would contribute $\sim$1 \% to 
the total luminosity of the multiple star. Therefore, the absence of any third light in our binary modeling does not rule out 
the existence of a circumbinary companion. 

The results presented in this work suggest that GV Leo is a potential triple system, (AB)C, that consists of a close binary 
(AB) with an eclipsing period of 0.2667 d and an outer, distant companion (C) with an LTT period of $\sim$15 years. The presence 
of the circumbinary companion may offer us important information on the origin and evolution of a tidal-locked close binary 
from a primordial widish binary by angular momentum and energy exchanges. This would have caused the eclipsing pair in GV Leo 
to evolve into its present contact state. When it meets the Darwin instability in which the sum of the spin angular momentum 
exceeds a third of the orbital one, the W UMa-type EB will coalesce into a single star \citep{darwin1879,hut1980}, so 
the potential triple system GV Leo will become a wide-orbit binary. This stellar merger is expected to result in a luminous 
red nova, as in the case of V1309 Sco \citep{tylenda2011}. 

The timing observations of GV Leo, spanning about 21 years, cover only 1.4 cycles of the 15-year LTT period. Hence, future 
high-precision eclipse measurements will help to verify our ETV analysis results for the system. Because our program target 
is relatively faint and the eclipsing pair has a short orbital period of 6.4 hr, 4-m class telescopes are required to measure 
its precise radial velocities (RVs). Combining the RV measurements with the astrometric data from Gaia and other facilities 
complement each other and greatly strengthen the astrophysical parameters of the contact binary, which should lead to 
a more detailed understanding of GV Leo's properties, such as its evolutionary status.


\acknowledgments

This paper is based on observations from LOAO, SOAO, and BOAO, which are operated by the Korea Astronomy and Space Science Institute 
(KASI). We wish to thank Dr. Kyeongsoo Hong for the spectroscopic observations of GV Leo and the anonymous referee for 
the careful reading and helpful comments. This research has made use of the Simbad database maintained at CDS, Strasbourg, 
France, and was supported by the KASI grant 2025-1-830-05. The works by M.-J.J. and C.-H. K. were supported by the grant numbers 
RS-2024-00452238 and 2021R1I1A3050979, respectively, from the National Research Foundation (NRF) of Korea.



\appendix
\section{List of Eclipse Timings}

In this appendix, we present historical CCD eclipse mid-times for GV Leo, together with our new measurements from 
the LOAO and SOAO observations. Here, $O$--$C_{\rm 2, full}$ represents the timing residuals from the full contribution 
of the quadratic {\it plus} LTT ephemeris, and Min I and II denote the primary and secondary minima, respectively. 
 
\renewcommand{\thetable}{\Alph{section}\arabic{table}}
\setcounter{table}{0}

\onecolumn
\begin{table}[ht!]
\centering
\caption{Observed CCD times of minima for GV Leo.\label{TabA1}}
{\begin{tabular}{llrccl}
\toprule
HJD           & Error         & Epoch       & $O-C_{\rm 2,full}$  & Min  & References                            \\ 
(2,400,000+)  &               &             &                     &      &                                       \\
\midrule
52,754.4598   & $\pm$0.0013   & $-$7725.0   & $-$0.00291          & I    & \citet{hubscher2005}                  \\
52,763.3966   & $\pm$0.0002   & $-$7691.5   & $-$0.00160          & II   & This paper \citep{frank2005}          \\
52,764.4639   & $\pm$0.0002   & $-$7687.5   & $-$0.00122          & II   & This paper \citep{frank2005}          \\
53,437.6973   & $\pm$0.0012   & $-$5163.5   & $+$0.00204          & II   & \citet{samec2006}                     \\
53,437.8293   & $\pm$0.0003   & $-$5163.0   & $+$0.00068          & I    & \citet{samec2006}                     \\
53,441.8291   & $\pm$0.0019   & $-$5148.0   & $-$0.00050          & I    & \citet{samec2006}                     \\
53,715.2308   & $\pm$0.0004   & $-$4123.0   & $+$0.00083          & I    & This paper (ASAS)                     \\
54,506.4949   & $\pm$0.0004   & $-$1156.5   & $+$0.00317          & II   & \citet{hubscher2010}                  \\
54,507.5613   & $\pm$0.0001   & $-$1152.5   & $+$0.00264          & II   & \citet{brat2008}                      \\
54,814.9668   & $\pm$0.0002   &       0.0   & $-$0.00068          & I    & \citet{nelson2009}                    \\
54,863.9128   & $\pm$0.0002   &     183.5   & $+$0.00002          & II   & \citet{diethelm2009}                  \\
54,900.7214   & $\pm$0.0002   &     321.5   & $-$0.00037          & II   & \citet{nelson2010}                    \\
54,908.4567   & $\pm$0.0001   &     350.5   & $-$0.00029          & II   & \citet{brat2009}                      \\
54,935.3964   & $\pm$0.0001   &     451.5   & $-$0.00049          & II   & \citet{brat2009}                      \\
55,243.7354   & $\pm$0.0004   &    1607.5   & $-$0.00285          & II   & \citet{diethelm2010}                  \\
55,289.3487   & $\pm$0.0008   &    1778.5   & $-$0.00051          & II   & \citet{hubscher2011}                  \\
55,289.4810   & $\pm$0.0010   &    1779.0   & $-$0.00158          & I    & \citet{hubscher2011}                  \\
55,589.8205   & $\pm$0.0008   &    2905.0   & $-$0.00027          & I    & \citet{diethelm2011}                  \\
55,589.9539   & $\pm$0.0003   &    2905.5   & $-$0.00023          & II   & \citet{diethelm2011}                  \\
55,597.2876   &               &    2933.0   & $-$0.00160          & I    & \citet{kriwattanawong2013}            \\
55,597.4217   &               &    2933.5   & $-$0.00086          & II   & \citet{kriwattanawong2013}            \\
55,598.2199   &               &    2936.5   & $-$0.00285          & II   & \citet{kriwattanawong2013}            \\
55,598.3543   &               &    2937.0   & $-$0.00182          & I    & \citet{kriwattanawong2013}            \\
55,599.2890   &               &    2940.5   & $-$0.00067          & II   & \citet{kriwattanawong2013}            \\
55,599.4238   &               &    2941.0   & $+$0.00077          & I    & \citet{kriwattanawong2013}            \\
55,599.5559   & $\pm$0.0005   &    2941.5   & $-$0.00050          & II   & \citet{gokay2012}                     \\
55,600.2203   &               &    2944.0   & $-$0.00292          & I    & \citet{kriwattanawong2013}            \\
55,600.3555   &               &    2944.5   & $-$0.00109          & II   & \citet{kriwattanawong2013}            \\
55,629.4297   & $\pm$0.0002   &    3053.5   & $-$0.00041          & II   & \citet{honkova2013}                   \\
55,671.7056   & $\pm$0.0006   &    3212.0   & $-$0.00112          & I    & \citet{diethelm2011}                  \\
55,674.3733   &               &    3222.0   & $-$0.00071          & I    & \citet{nagai2012}                     \\
55,676.3749   &               &    3229.5   & $+$0.00042          & II   & \citet{nagai2012}                     \\
55,678.3747   &               &    3237.0   & $-$0.00025          & I    & \citet{nagai2012}                     \\
55,953.9083   & $\pm$0.0001   &    4270.0   & $+$0.00253          & I    & \citet{diethelm2012}                  \\
55,957.6424   & $\pm$0.0002   &    4284.0   & $+$0.00244          & I    & \citet{honkova2013}                   \\
55,963.3752   & $\pm$0.0005   &    4305.5   & $+$0.00058          & II   & \citet{honkova2013}                   \\
56,014.3221   & $\pm$0.0002   &    4496.5   & $+$0.00242          & II   & \cite{gursoytrak2013}                 \\
56,017.6556   & $\pm$0.0002   &    4509.0   & $+$0.00182          & I    & \citet{diethelm2012}                  \\
56,246.6410   & $\pm$0.0002   &    5367.5   & $+$0.00167          & II   & \citet{honkova2013}                   \\
56,304.2546   &               &    5583.5   & $+$0.00225          & II   & \citet{nagai2014}                     \\
56,339.9969   &               &    5717.5   & $+$0.00317          & II   & \citet{nagai2014}                     \\
56,340.1287   &               &    5718.0   & $+$0.00160          & I    & \citet{nagai2014}                     \\
56,340.2631   &               &    5718.5   & $+$0.00264          & II   & \citet{nagai2014}                     \\
56,630.5903   & $\pm$0.0003   &    6807.0   & $-$0.00158          & I    & \citet{honkova2015}                   \\
56,685.1377   &               &    7011.5   & $+$0.00045          & II   & \citet{nagai2015}                     \\
56,685.2690   &               &    7012.0   & $-$0.00161          & I    & \citet{nagai2015}                     \\
56,716.3497   & $\pm$0.0014   &    7128.5   & $+$0.00558          & II   & \citet{honkova2015}                   \\
57,067.4870   & $\pm$0.0003   &    8445.0   & $-$0.00042          & I    & \citet{jurysek2017}                   \\
57,067.6190   & $\pm$0.0002   &    8445.5   & $-$0.00178          & II   & \citet{jurysek2017}                   \\
57,096.0293   &               &    8552.0   & $+$0.00237          & I    & \citet{nagai2016}                     \\
57,102.6941   & $\pm$0.0001   &    8577.0   & $-$0.00094          & I    & \citet{samolyk2016}                   \\
57,105.3620   & $\pm$0.0001   &    8587.0   & $-$0.00028          & I    & \citet{jurysek2017}                   \\
\bottomrule
\end{tabular}
}
\end{table}
\twocolumn
\setcounter{table}{0}
\onecolumn
\begin{table}[ht!]
\centering
\caption{Continued.}
{\begin{tabular}{llrccl}
\toprule
HJD           & Error         & Epoch       & $O-C_{\rm 2,full}$  & Min  & References                            \\ 
(2,400,000+)  &               &             &                     &      &                                       \\
\midrule
57,121.7674   & $\pm$0.0001   &    8648.5   & $+$0.00157          & II   & \citet{samolyk2016}                   \\
57,400.0923   &               &    9692.0   & $-$0.00011          & I    & \citet{nagai2017}                     \\
57,457.7023   & $\pm$0.0001   &    9908.0   & $-$0.00247          & I    & \citet{samolyk2017}                   \\
57,800.0440   &               &   11191.5   & $-$0.00067          & II   & \citet{nagai2018}                     \\
57,800.1780   &               &   11192.0   & $-$0.00003          & I    & \citet{nagai2018}                     \\
57,800.3107   &               &   11192.5   & $-$0.00070          & II   & \citet{nagai2018}                     \\
57,822.1824   & $\pm$0.0004   &   11274.5   & $-$0.00034          & II   & This paper (SOAO)                     \\
57,854.9896   & $\pm$0.0002   &   11397.5   & $-$0.00015          & II   & This paper (SOAO)                     \\
58,141.84992  & $\pm$0.00007  &   12473.0   & $-$0.00123          & I    & This paper (LOAO)                     \\
58,141.98390  & $\pm$0.00008  &   12473.5   & $-$0.00062          & II   & This paper (LOAO)                     \\
58,142.91639  & $\pm$0.00006  &   12477.0   & $-$0.00166          & I    & This paper (LOAO)                     \\
58,143.85111  & $\pm$0.00009  &   12480.5   & $-$0.00047          & II   & This paper (LOAO)                     \\
58,143.98368  & $\pm$0.00007  &   12481.0   & $-$0.00126          & I    & This paper (LOAO)                     \\
58,145.85088  & $\pm$0.00009  &   12488.0   & $-$0.00113          & I    & This paper (LOAO)                     \\
58,145.98488  & $\pm$0.00007  &   12488.5   & $-$0.00049          & II   & This paper (LOAO)                     \\
58,157.1873   & $\pm$0.0005   &   12530.5   & $-$0.00048          & II   & This paper (SOAO)                     \\
58,158.2546   & $\pm$0.0005   &   12534.5   & $-$0.00007          & II   & This paper (SOAO)                     \\
58,187.1930   & $\pm$0.0002   &   12643.0   & $-$0.00121          & I    & This paper (SOAO)                     \\
58,460.3186   & $\pm$0.0001   &   13667.0   & $-$0.00103          & I    & This paper (SOAO)                     \\
58,462.3200   & $\pm$0.0001   &   13674.5   & $-$0.00007          & II   & This paper (SOAO)                     \\
58,914.15299  & $\pm$0.00004  &   15368.5   & $+$0.00158          & II   & This paper (SOAO)                     \\
59,198.2163   & $\pm$0.0002   &   16433.5   & $+$0.00276          & II   & This paper (SOAO)                     \\
59,212.2153   & $\pm$0.0002   &   16486.0   & $-$0.00132          & I    & This paper (SOAO)                     \\
59,235.1540   & $\pm$0.0001   &   16572.0   & $-$0.00098          & I    & This paper (SOAO)                     \\
59,251.1572   & $\pm$0.0002   &   16632.0   & $-$0.00130          & I    & This paper (SOAO)                     \\
59,279.0338   & $\pm$0.0001   &   16736.5   & $+$0.00252          & II   & This paper (SOAO)                     \\
59,304.1053   & $\pm$0.0001   &   16830.5   & $+$0.00185          & II   & This paper (SOAO)                     \\
59,634.0417   & $\pm$0.0002   &   18067.5   & $-$0.00100          & II   & This paper (SOAO)                     \\
59,634.17466  & $\pm$0.00007  &   18068.0   & $-$0.00141          & I    & This paper (SOAO)                     \\
60,006.1232   & $\pm$0.0001   &   19462.5   & $-$0.00131          & II   & This paper (SOAO)                     \\
60,402.07859  & $\pm$0.00004  &   20947.0   & $+$0.00105          & I    & This paper (SOAO)                     \\
60,434.08553  & $\pm$0.00004  &   21067.0   & $+$0.00106          & I    & This paper (SOAO)                     \\
\bottomrule
\end{tabular}
}
\end{table}

\end{document}